\documentstyle[floats,twocolumn,aps,prl]{revtex}
\begin{document}
\ifpreprintsty\else
\twocolumn[\hsize\textwidth%
\columnwidth\hsize\csname@twocolumnfalse\endcsname
\fi
\title{Magnetic field induced luminescence spectra in a quantum cascade laser}
\author{V.M. Apalkov and T. Chakraborty}
\address{Max-Planck-Institut f\"ur Physik Komplexer Systeme,
Dresden, Germany}
\maketitle
\begin{abstract}
We report on our study of the luminescence spectra of a quantum cascade
laser in the presence of an external magnetic field tilted from the
direction perpendicular to the electron plane. The effect of the tilted
field is to allow novel optical transitions because of the coupling of
intersubband-cyclotron energies. We find that by tuning the applied
field, one can get optical transitions at different energies that are
as sharp as the zero-field transitions.
\end{abstract}
\ifpreprintsty\clearpage\else\vskip1pc]\fi
%\pacs{42.55.Px,76.60.Jx,73.40.Gk,78.60,-b}
\narrowtext

The unipolar quantum cascade laser (QCL) 
\cite{first,vertical,nature} is the product of ingenious quantum 
engineering that exploits the properties of electrons confined in 
semiconductor nanostructures. As yet, this is the only high power 
semiconductor laser that operates at and above room temperature in 
the mid-infrared range 
\cite{shortwave,longwave,room,aboveroom,gaas1,gaas2}. Intense 
interest on this system derives from its technological importance 
in trace-gas analysis, in  particular, for environmental control, 
remote chemical sensing, pollution monitoring, non-invasive medical 
diagnostics, etc. \cite{aboveroom}. Here we study the novel effects
of a tilted magnetic field on a QCL where the coupled
intersubband-cyclotron transitions are allowed \cite{book,tilt}. We 
find that as the subbands quantize into discrete Landau levels, new 
luminescence peaks appear that correspond to those transitions. 
The peaks exhibit a prominent red shift but can be made as 
sharp and large as the zero-field case by tuning the applied field.

The system we have studied here is sketched schematically in Fig. 1,
where a GaInAs quantum well of 7.4 nm width is sandwiched between two
AlInAs tunneling barriers. When a suitable bias is applied, electrons
tunneling through the upstream barrier generate photons and escape
quickly to the next well through the downstream barrier. Superlattice
structures on both sides of the active region act as electron injector
or Bragg mirrors and control the rate of electron escape 
\cite{first,vertical,nature}. The operating wavelength of the lasers
is determined by the quantum confinement rather than the bandgap
of the materials.

\begin{figure}
\begin{center}
\begin{picture}(120,130)
\put(0,0){\includegraphics{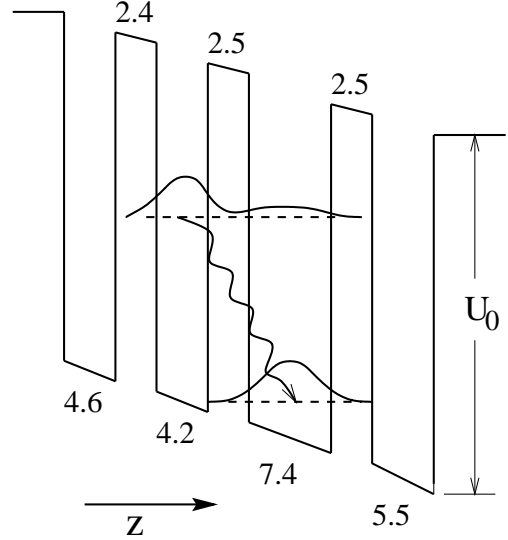}}
\end{picture}
\vspace*{-1.0cm}
\caption{Energy band diagram (schematic) of the active region of
a quantum cascade laser structure under an average applied electric 
field of 55 kV/cm. Only one period of the device is shown here. 
The relevant wave functions (moduli squared) as well as
the transition corresponding to the laser action are also shown
schematically. The numbers (in nm) are the well (Ga$_{0.47}$In$_{0.53}$As)
and barrier (Al$_{0.48}$In$_{0.52}$As) widths. 
Material parameters considered in this work are: electron 
effective mass $m_e^*$ (Ga$_{0.47}$In$_{0.53}$As)=0.043 $m_0$, 
$m_e^*$ (Al$_{0.48}$In$_{0.52}$As)=0.078 $m_0$, the conduction band
discontinuity, $U_0=520$ meV, the nonparabolicity coefficient,
$\gamma_w=1.3\times10^{-18}$ m$^2$ for the well and 
$\gamma_b=0.39\times10^{-18}$ m$^2$ for the barrier, and the sheet
carrier density induced by doping, $n_s=2.3\times10^{11}$
cm$^{-2}$. The energy difference between the two levels where the 
optical transition takes place, is 132 meV. All computations were 
performed at $T=50$ K.}
\end{center}
\end{figure}

An externally applied magnetic field tilted from the
direction perpendicular to the electron plane is a well-studied 
problem experimentally as well as theoretically, in the context of 
quantum Hall effects \cite{book}. For a magnetic field tilted
from the $z$ direction the perpendicular and parallel motions of
electrons are coupled and as a result, transitions between different
Landau levels of the ground and upper subbands become possible.
Interestingly, the perpendicular component of the tilted field provides 
magnetic quantization, somewhat analogous to the situation proposed for 
a quantum-dot cascade laser \cite{box,dot} where the quantization of 
the planar motion is due to replacement of the quantum wells by quantum 
dots \cite{qdbook}. The Hamiltonian of the system in a tilted field is 
$${\cal H} = {\cal H}_{\perp} + {\cal H}_{\parallel} + 
{\cal H}^{\prime}$$
where
\begin{eqnarray*}
{\cal H}_\perp &=& \frac1{2m^*}p_z^2 + V_{\rm eff}(z) + 
\frac{\hbar^2}{2m^*}\frac{z^2}{\ell_{\parallel}^4}\\
{\cal H}_\parallel &=& \frac1{2m^*}\left[p_x^2 + \frac{\hbar^2}{
\ell_\perp^4}(x+X)^2\right]\\
{\cal H}^{\prime } &=& \frac{\hbar}{m^*}
\frac{zp_x}{\ell_{\parallel}^2}\\
\end{eqnarray*}
with $X\equiv \frac{p_y}{\hbar}\ell_\perp^2$ the center coordinate of
the cyclotron motion and $\ell_\parallel^2=c\hbar/eH_y$, 
$\ell_\perp^2=c\hbar/eH_z$ are the magnetic lengths. Here we have 
chosen the Landau gauge vector potential, ${\bf A}=(H_yz, H_zx,0)$. 
The magnetic field is therefore in the $y-z$ plane and $H_y=H\sin\theta, 
H_z=H\cos\theta$, $\theta$ is the tilt angle and $H$ is the total 
magnetic field \cite{book}.

The effective potential is made up of the (i) confinement
potential, (ii) Hartree potential, and the (iii) exchange-correlation 
potential \cite{alder}. The wave functions of the Hamiltonian 
${\cal H}_\perp$, which depend only on $z$-coordinate are obtained from
$${\cal H}_\perp\psi_n = E_n \psi_n(z).$$
Solutions of this equation determine the energy levels and wave 
functions of the subbands. The total Hamiltonian is diagonalized by 
choosing the basis wave functions
\begin{eqnarray*}
\Psi_{n,N,X} &=& L^{-\frac12}\exp\left(-{\rm i}\frac Xy 
{\ell_\perp^2}-{\rm i} \frac{z_{nn}}{\ell_\parallel^2} (x-X)\right)\\
&\times&\xi_{N}(x-X)\psi_n(z)\\
\xi_N(x) &=& {\rm i}^N (2^N N! \pi^{\frac12}\ell_\perp)^{-1/2} H_{N}\left(
\frac x{\ell_\perp}\right) \exp\left(-\frac{x^2}{2\ell_\perp^2}\right)\\
z_{nm} &=& \int dz\, \psi_n(z)\,z\,\psi_m(z)\\
\end{eqnarray*}
where $H_N(x)$  is the Hermite polinomial. The Hamiltonian matrix elements
$(n^{\prime}N^{\prime }X^{\prime }|{\cal H}|nNX)$
are then calculated in this basis. The matrix is diagonal in $X$. 
To calculate the wave functions we use three subbands and 20 Landau 
lebels on each subband.

Once we diagonalize the Hamiltonian the chemical potential is 
then obtained from the equation
$$N_s = \frac2{2\pi\ell_\perp^2}\sum_N \frac1{\exp(\beta(E_{2,N}
-\mu))+1}$$
where $E_{2,N}$ is the energy of $N$-th Landau level in the second
subband (only the second subband is occupied). The optical spectra 
are calculated as follows \cite{tilt}: Optical transitions occur 
between Landau levels of the second and first subbands. We introduce 
the variables that are proportional to matrix elements of optical 
transitions
\begin{eqnarray*}
u_{(N,N_1)} &=& \frac1{eF}\left(\frac{2m^*}{\hbar^2}
\right)^{-\frac12}\frac{(1,N|{\cal H}^\ast|2,N_1)}{(E_{2,N_1}
-E_{1,N})^2-(\hbar\omega)^2}\\
&\times&\left\{f(E_{2,N_1})\right\}^{\frac12}
\end{eqnarray*}
where ${\cal H}^\ast$ is the change of Hamiltonian due to an 
external electric field $F$ in the $z$ direction, and
$f(\varepsilon)$ is the Fermi distribution function. Note that
we consider the first subband as empty. The emission spectrum is  
related to $u_{(N,N_1)}$ through 
$${\cal I}(\omega) \propto \omega \frac1{2\pi\ell_\perp^2}
\sum_{N,N_1} z_{1,2}\left\{f(E_{2,N_1})\right\}^{\frac12}
\Im(u_{(N,N_1)})$$
where $\omega$ is the frequency of the emitted photon.

\begin{figure}
\begin{center}
\begin{picture}(120,130)
\put(0,0){\includegraphics{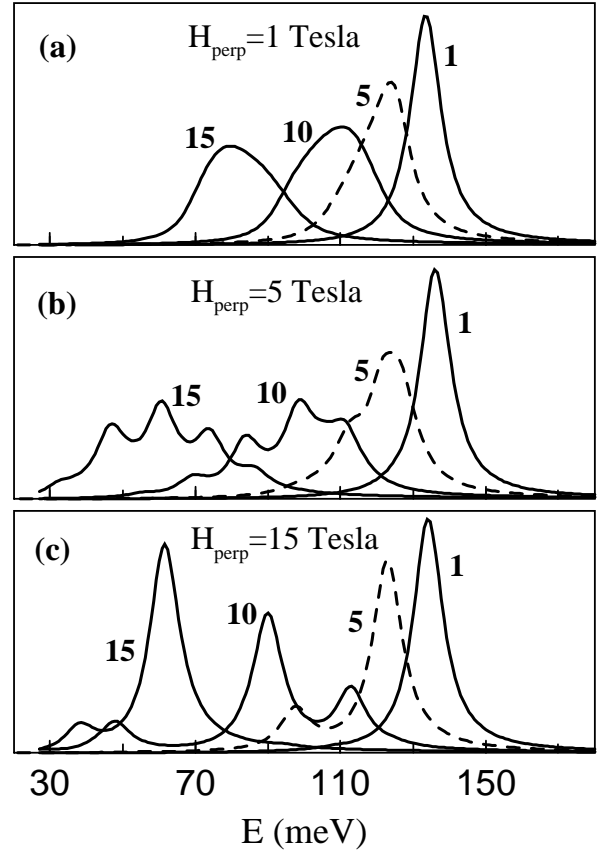}}
\end{picture}
\vspace*{7.0cm}
\caption{Luminescence spectra at various values of the parallel
component of the magnetic field (numbers by the curves in tesla)
for a fixed value of the perpendicular component of the field.}
\end{center}
\end{figure}

Intensity of the optical emission is determined by the dipole matrix
elements between initial (before emission) and final (after emission)
states of the multi-electron system. In dipole transitions the
transition intensity is proportional to the overlap between 
$(x,y)$-dependent parts of the wave functions of the initial and final 
states \cite{qdbook}. For a vanishing perpendicular magnetic field this 
results in conservation of the two-dimensional momentum in the optical 
transition. For non-zero perpendicular magnetic fields the states 
of the two-dimensional
electrons are classified by two numbers: the Landau level index
and by a number that distinguishes the degenerate states of an electron 
within a Landau level, for example, by $x$ component of the 
momentum. The energy of the single electron system depends only on the 
number of the Landau level, and the wave functions are harmonic
oscillator functions whose center is determined by the
$x$ component of the momentum. If the magnetic field is directed 
perpendicular to the two-dimensional layer then optical transitions 
are allowed only between the states with the same two-dimensional
quantum numbers. In this case we will have a single line which 
corresponds to the optical transitions between states with the same 
Landau level index. 

Introduction of a non-zero parallel magnetic field in $y$ direction 
results in modification of the wave function in a Landau level: the 
position of oscillator wave functions is now determined by 
$y$ component of the momentum and also by average position of the 
electron in $z$ direction, $\langle z\rangle$ that depends on the 
number of subbands. Figure 1 illustrate this dependence. Electrons 
in the first subband and in the second subband are localized in the 
quantum wells 7.4 nm and 4.2 nm, respectively. This opens up the 
possibility for optical transitions between different Landau levels 
and at the same time it suppresses the optical transitions between 
the states with the same Landau level indices.   

Evolution of the emission spectra as a function of the tilted 
magnetic field and tilt angle are illustrated in Figs. 2 and 3. In 
Fig. 2, the optical spectra are shown for three values of the 
perpendicular component of the field and for different values of 
parallel component of field for a fixed $H_{\rm perp}$. Clearly, for 
small $H_{\rm perp}$ (Fig. 2a), the emission spectra do not feel the 
Landau quantization and we have a single peak which broadens with 
increasing $H_{\rm par}$. For higher fields such as $H_{\rm perp}=5$ 
tesla (Fig.2b), we find new features in the emission 
spectra. For a small parallel field, $H_{\rm par}=1$ tesla, main 
transitions are between the states with the same Landau index,
hence a single peak that corresponds to transitions from the zeroth 
Landau level of second subband to that of the first subband. 
An increase of the parallel field makes transitions to higher Landau 
levels more intense. Appearance of a shoulder at $H_{\rm par}=5$ tesla 
corresponds to transitions to the first Landau level of the first 
subband. Similarly, peaks at $H_{\rm par}=10$ tesla and 15 tesla 
correspond to transitions from zeroth Landau level of the second 
subband to the higher Landau level of the first subband. The energy 
separations between the peaks are equal to the separations between 
Landau levels of the first subband. In Fig. 2c, transitions to 
non-zero Landau level become more intense and we observe an 
interplay between the transitions to zero and to the first Landau 
levels with increasing parallel field. For a small $H_{\rm par}$ 
($H_{\rm par}=1$ tesla), there is only a strong transition to zeroth 
Landau level. For $H_{\rm par}=5$ tesla, we observe the appearance of 
a small peak that corresponds to transitions to the first Landau level, 
and for $H_{\rm par}=10$ tesla and 15 tesla, transitions to the first 
Landau level become strongest and we observe the formation of 
a new narrow peak that corresponds to a transition to the first 
Landau level.

\begin{figure}
\begin{center}
\begin{picture}(120,130)
\put(0,0){\includegraphics{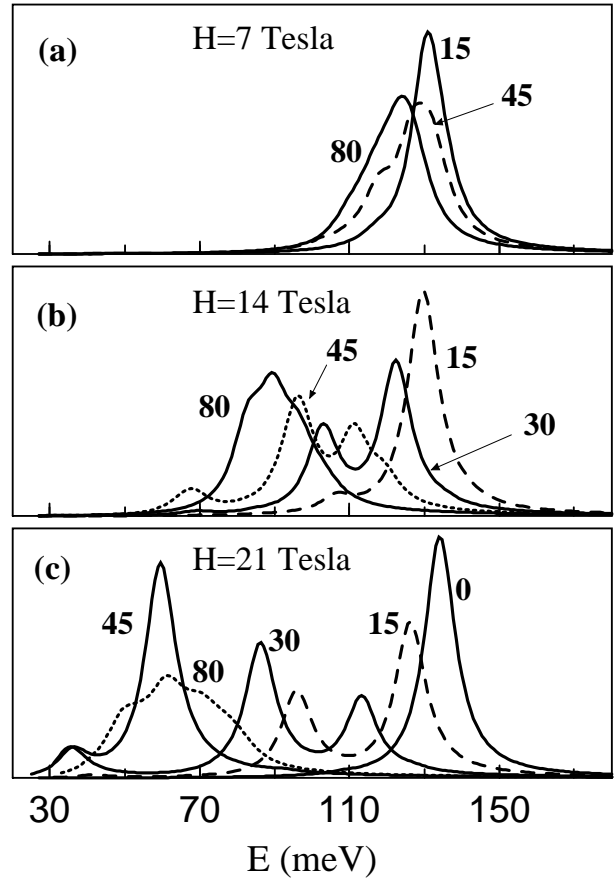}}
\end{picture}
\vspace*{8.0cm}
\caption{Luminescence spectra at various values of the tilt angle 
(numbers by the curves) for a fixed value of the total magnetic field. 
}
\end{center}
\end{figure}

In a recent experiment by Blaser et al. \cite{blaser}, QCL based on
photon-assisted tunneling transition was subjected to a parallel
field of upto 14 tesla. The luminescence spectra showed a
broadening of the peak but almost no shift (a very tiny blue shift 
was observed) with an increasing parallel field. Broadening of the 
luminescence peak with increasing parallel field is also visible in 
Fig. 2(a) where the perpendicular field is rather small, but the 
experimentally observed broadening is much more rapid. In addition, 
there is a strong red shift in Fig. 2(a) that is not visible in the 
observed spectra. In the present scheme, a red shift seems to
be quite natural in the single-electron picture \cite{ando,partheo}.
The change of the wave function due to a parallel field 
($H_{\rm par}<10$ tesla) is small for the first and the second 
subbands. Magnetic field changes the energy spectrum because of a 
shift of the parabolic energy dispersion by a value that is 
proportional to the field. This results in a broadening and red shift 
of the emission spectrum without the many-body effects. The many-body 
effects changes the results only slightly and results in a 
blue shift ($\sim$ 4 meV) of the emission spectra that is independent
of the applied magnetic field. The observed total 
intensity of emission line gets smaller with higher parallel fields.
At 10 tesla, it is smaller than its value at zero field ($I_0$) by 
45\% (0.55 $I_0$) \cite{blaser2}. In contrast, our calculated values 
indicate a reduction by 16\% (0.84 $I_0$).

In our present scheme, the small blue shift (or almost absence of
any shift) observed in \cite{blaser} in a strong parallel field cannot 
be explained by many-body corrections, because these corrections 
in samples of \cite{blaser} are small and have a weak magnetic field 
dependence. The weak magnetic field dependence of many-body effects 
was also demonstrated in \cite{meester} for much wider quantum wells 
(30-46 nm) where one expects even larger magnetic field dependence
than for narrow wells. One possible explanation of the observed results 
of \cite{blaser} might be due to the presence of disorder in the system
which tends to localize the two-dimensional electrons \cite{metzner} and 
thereby break the conservation of momentum in the optical emission. 
Magnetic field tends to delocalize the states which can result 
in a decrease of emission intensity and widening of the emission 
lines. Our present results would, of course, be valid for 
disorder-free high-mobility systems. 

In Fig. 3, the emission spectra are shown for three values of the
total magnetic field $H$ and for different values of the tilt angle
at a fixed $H$. For a small field (Fig. 3a) we have a red shift of the
emission spectra with increasing parallel field (i.e., increasing tilt
angle). For $\theta=45^\circ$ there is a weak structure resulting from 
the Landau quantization. At higher fields (Fig. 3b,c), one observes
the evolution of the emission spectra from a broad peak at a large angle, 
$\theta=80^\circ$, (large parallel field and a small perpendicular 
field) to a single narrow peak for small angle. In the latter case,
the parallel component of the magnetic field is small and all 
optical transitions are transitions between the Landau levels
with the same index. For an intermediate tilt angle, there are two peaks
that correspond to transitions from the zeroth Landau level of 
the second subbband to zeroth and the first Landau levels of the 
first subband. The intensity of transition to the first Landau level 
increases  with increasing angle which means an increase of the
parallel field. It has its maximum at $\theta=45^\circ$ and for total 
magnetic field of $H=21$ tesla one observes the formation of a strong 
narrow peak at  $\theta=45^\circ$ associated with a suppression of the 
original peak corresponding to the transition to zeroth Landau level 
of the first subband. This is the same peak as shown in Fig. 2 for 
$H_{\rm perp}=H_{\rm par}=15$ tesla. From these results, we conclude 
that by suitably tuning the externally applied tilted field, lasing 
due to coupled intersubband-cyclotron transitions that is as strong 
as the zero-field case (but at different energies) can be acheived. 
Although for the present system the magnetic fields where the novel 
transitions take place are rather high, by suitably engineering the 
device parameters these transitions can be made to occur at lower fields.

We would like to express our gratitude to P. Fulde for his kind support
and for valuable discussions. We also thank S. Blaser for helpful 
discussions and for his unpublished results.

\vfil
\end{document}